\documentclass[
a4paper,
aps,
prd,
preprint,
superscriptaddress,
showpacs,
showkeys,
nofootinbib
]{revtex4}

\usepackage{enumitem}
\usepackage{pifont}
\usepackage{fullpage}
\usepackage{amsfonts}
\usepackage{amsmath}
\usepackage{amssymb} 
\usepackage{graphicx}
\usepackage{epic}
\usepackage{eepic}
\usepackage{epsfig}
\usepackage{latexsym}
\usepackage{color}
\usepackage{float}
\usepackage{multirow} 
\usepackage[citecolor=green]{hyperref} 
\usepackage[caption=false]{subfig}

\usepackage{shorthand}

\begin{document}

\title{Effect of strong magnetic field on competing order parameters in two-flavor dense quark matter}

\author{Tanumoy Mandal}
\email{tanumoy.mandal@physics.uu.se}
\affiliation{Department of Physics and Astronomy, Uppsala University, Box 516, SE-751 20 Uppsala, Sweden}

\author{Prashanth Jaikumar}
\email{prashanth.jaikumar@csulb.edu}
\affiliation{Department of Physics and Astronomy, California State University Long Beach, Long Beach, CA 90840 USA}

\date{\today}

\begin{abstract} 
We study the effect of strong magnetic field on competing chiral and diquark order parameters in a
regime of moderately dense quark matter. The inter-dependence of the chiral and diquark condensates 
through nonperturbative quark mass and strong coupling effects is analyzed in a two-flavor 
Nambu-Jona-Lasinio (NJL) model. In the weak magnetic field limit, our results agree qualitatively 
with earlier zero-field studies in the literature that find a critical coupling ratio $G_D/G_S\sim 1.1$
below which chiral or superconducting order parameters appear almost exclusively. Above the critical
ratio, there exists a significant mixed broken phase region where both gaps are non-zero. However, 
a strong magnetic field $B\gtrsim 10^{18}$~G disrupts this mixed broken phase region and changes a
smooth crossover found in the weak-field case to a first-order transition for both gaps at almost 
the same critical density. Our results suggest that in the two-flavor approximation to moderately
dense quark matter, strong magnetic field enhances the possibility of a mixed phase at high density,
with implications for the structure, energetics and vibrational spectrum of neutron stars.
\end{abstract}



\maketitle
 
\section{Introduction}
\label{sec_intro}

The existence of deconfined quark matter in the dense interior of a neutron star is 
an interesting question that has spurred research in several new directions in nuclear 
astrophysics. On the theoretical side, it has been realized that cold and dense quark 
matter must be in a superconductor/superfluid state
\cite{Barrois:1977xd,Bailin:1983bm,Iwasaki:1994ij,Alford:1997zt,Rapp:1997zu,Alford:2007xm}
with many possible intervening phases~\cite{Alford:1998mk,Huang:2001yw,Mishra:2004gw,Steiner:2002gx,Neumann:2002jm,Shovkovy:2003uu,Schmitt:2004hg,Rajagopal:2006ig} between a few times nuclear matter density to
asymptotically high density, where quarks and gluons interact weakly. 
The observational impact of these phases
on neutron star properties can be varied and 
dramatic~\cite{Alcock:1986hz,Glendenning:1997fy,Page:2002bj,Jaikumar:2002vg,Reddy:2002xc,Jaikumar:2005hy,Berdermann:2016mwt,
Jaikumar:2008kh}. 
Therefore, it is of interest to situate theoretical ideas and advances 
in our understanding of dense quark matter in the context of neutron
stars, which serve as unique astrophysical laboratories for
such efforts. The phase structures of hot quark matter have been probed in experiments such
as in heavy-ion collisions at the Relativistic Heavy Ion Collider (RHIC) and at the Large
Hadron Collider (LHC). It is estimated in Refs.~\cite{Kharzeev:2007jp,Skokov:2009qp,Voronyuk:2011jd} that the
magnetic field originating from off-central nucleon-nucleon collisions at these colliders can 
be as large as $10^{18}-10^{20}$ G. On the astrophysical side, the strength
of the magnetic field in some magnetars is of the order $10^{14}-10^{15}$ G~\cite{Paczynski:1992zz}, while in the core of such objects, magnetic field might reach up to $10^{18}-10^{19}$ G. Therefore, it is not surprising that many recent works have stressed the role of strong magnetic fields on hot or dense quark matter~\cite{Mandal:2012fq,
Allen:2013lda,Sinha:2013dfa,Denke:2013gha,Grunfeld:2014qfa,Chatterjee:2014qsa,Allen:2015qxa,Denke:2015kga,Mandal:2016dzg}. 

At very high density ({\it i.e.} $\mu\gg \Lambda_{\rm QCD}$ where $\mu$ is the 
baryon chemical potential and $\Lm_{\textrm{QCD}}$ is the scale of quantum chromodynamics) and 
for number of flavors $N_f=3$, 
the preferred pairing pattern is a flavor and color democratic one termed as the 
color-flavor-locked (CFL) phase~\cite{Alford:1998mk}. This idealized phase, while it displays the essentially novel features of the color superconducting state,
is unlikely to apply to the bulk of the neutron star matter, since even ten times 
nuclear matter saturation density ($\rho_0$) only corresponds to a quark chemical potential $\mu\sim 500$ MeV. At these densities, quark mass and strong coupling effects can be important,
and must be treated nonperturbatively. It is reasonable to think that the strange quark current mass,
being much larger than that of the up and down quarks, inhibits pairing of strange quarks with light
quarks. For the purpose of this work, we therefore adopt the scenario of quark matter in the two-flavor superconducting phase, which breaks the color $\mathrm{SU}(3)$ symmetry to $\mathrm{SU}(2)$, leaving light quarks
of one color (say ``3'') and all colors of the strange quark unpaired. Although this phase initially
appeared to be disfavored in compact stars~\cite{Alford:2002kj,Aguilera:2004ag} once constraints of neutrality were imposed within a perturbative approach to quark masses, the NJL model where masses are treated dynamically still allows for the 2SC phase. Since the issue is not settled, we proceed by adopting the NJL model which best highlights the competition between the chiral and diquark condensates in a straightforward way. Also, our results will be qualitatively true for the 2SC+s phase~\cite{Steiner:2002gx,Mishra:2004gw}, which can be studied similarly by simply embedding the strange quark, which is inert with respect to pairing, in the enlarged three-flavor space. The additional complications of compact star constraints have been examined 
before~\cite{Fayazbakhsh:2010gc,Mandal:2012fq,Mandal:2016dzg}, and do not change the main qualitative conclusions of the present work, namely, that strong magnetic field alters the competition between the chiral and diquark order parameters from the weak-field case.

Our objective in this paper is a numerical study of the competition between the chiral 
and diquark condensates at moderately large $\mu$ and large magnetic field using 
the NJL model, similar in some respects to previous works
\cite{Huang:2001yw,Schwarz:1999dj,Buballa:2001gj,Mishra:2004gw,Mishra:2003nr}, 
which treat the quark mass non-perturbatively. Instanton-based calculations and 
random-matrix methods have also been employed in studying the interplay of 
condensates~\cite{Berges:1998rc,Carter:1999xb,Vanderheyden:1999xp}. In essence, 
smearing of the Fermi surface by diquark pairing can affect the onset of chiral 
symmetry restoration, which happens at $\mu\sim M_q$,
where $M_q$ is the constituent quark mass scale~\cite{Chen:2008zr}. Since $M_q$ appears also 
in the (Nambu-Gorkov) quark propagators in the gap equations, a coupled analysis of chiral 
and diquark condensates is required. This was done for the two-flavor case with a common 
chemical potential in~\cite{Huang:2001yw}, but for zero magnetic field. We use a self-consistent 
approach to calculate the condensates from the coupled gap equations, and find small 
quantitative (but not qualitative) differences from the results of Huang et. al.~\cite{Huang:2001yw} 
for zero magnetic field. This small difference is most likely attributed to a difference in numerical 
procedures in solving the gap equations. We also address the physics of chiral and diquark 
condensates affected by large in-medium magnetic field that are generated by circulating 
currents in the core of a neutron or hybrid star. Magnetic field in the interior of neutron stars may be as large as $10^{19}$~G, 
pushing the limits of structural stability of the star~\cite{Bocquet:1995je,Broderick:2001qw}. 
There is no Meissner effect for the rotated photon, which has only a small gluonic component, 
therefore, magnetic flux is hardly screened~\cite{Alford:1999pb}, implying that studies of 
magnetic effects in color superconductivity are highly relevant. Note that the rotated gluonic field,
which has a very small photonic component, is essentially screened due to the 2SC phase.
Including the magnetic interaction of the quarks with the external field leads to qualitatively 
different features in the competition between the two condensates, and this is the main result of our work. 

In Section~\ref{sec:lag}, we state the NJL model Lagrangian for the 2SC quark matter.
In Section~\ref{sec:therpot}, we recast the partition function and thermodynamic potential in terms of interpolating bosonic variables. In Section~\ref{sec:gapeqn}, we obtain the gap equations for the chiral and diquark order parameters by minimizing the thermodynamic potential (we work at zero temperature throughout since typical temperature in stars, $T_{\rm star}\ll \mu$). In Section~\ref{sec:nuana}, we discuss our numerical results for the coupled evolution of the condensates as  functions of a single ratio of couplings, chemical potential and magnetic field before concluding in Section \ref{sec:conclu}. 

\section{Lagrangian for 2SC quark matter}
\label{sec:lag}

The Lagrangian density for two quark flavors ($N_f=2$) applicable to the 
scalar and pseudoscalar mesons and scalar diquarks is
\begin{eqnarray}
\label{eq:NJLlag}
\mathcal{L} &=& \bar{q}\left[i\gamma^{\mu}\left(\partial_{\mu} - ieQA_{\mu} - igT^8G^8_{\mu}\right) 
+ \hat{\mu}\gamma^{0} - \hat{m}\right]q + G_{S}\left[\left(\bar{q}q\right)^2
+ \left(\bar{q}i\gamma_{5}\vec{\tau}q\right)^2\right]\nonumber\\
&+& G_D\left[\left(\bar{q}i\gamma_{5}\epsilon_{f}\epsilon_{c}q^C\right)
\left(\bar{q}^{C}i\gamma_{5}\epsilon_{f}\epsilon_{c}q\right)\right]\ ,
\end{eqnarray}
where $q\equiv q_{ia}$ is a Dirac spinor which is a doublet (where $i=\{u,d\}$) in
flavor space and triplet (where $a=\{1,2,3\}$) in color space. The charge-conjugated
fields are defined as $\bar{q}^C=-q^TC$ and $q^C=C\bar{q}^T$ with 
charge-conjugation matrix $C=-i\gamma^0\gamma^2$. The components of the 
$\vec{\tau}=(\tau^1,\tau^2,\tau^3)$ are the Pauli matrices in flavor space and, 
$(\epsilon_f)_{ij}$ and $(\epsilon_c)^{\alpha\beta 3}$ are the antisymmetric matrices 
in flavor and color spaces respectively. The common quark chemical potential
is denoted as $\hat{\mu}$
\footnote{For simplicity we assume a common chemical potential for all quarks. In an actual 
neutron star containing some fraction of charge neutral 2SC or 2SC+s 
quark matter in $\beta$-equilibrium, additional chemical potentials for electric 
charge and color charges must be introduced in the NJL model. Furthermore,
there can be more than one diquark condensate and in general 
$M_u\neq M_d\neq M_s$~\cite{Mishra:2004gw}.}
and $\hat{m}={\rm diag}(m_{u},m_{d})$ is the current quark mass matrix in the flavor 
basis. We take the exact isospin symmetry limit, $m_{u}=m_{d}=m_{0}\neq 0$.
The $\mathrm{U}(1)$ and $\mathrm{SU}(3)_c$ gauge fields are denoted by $A_{\mu}$ and $G_{\mu}$ respectively.
Here, $e$ is the electromagnetic charge of an electron and $g$ is 
the $\mathrm{SU}(3)_c$ coupling constant.  
The electromagnetic charge matrix for quark is defined as $Q=Q_f\otimes \bf 1_c$ 
with $Q_f\equiv\textrm{diag}(2/3,-1/3)$ (in unit of $e$). The couplings of the
scalar and diquark channels are denoted as $G_S$ and $G_D$ respectively.
In general, one can extend the NJL Lagrangian considered in Eq.~\ref{eq:NJLlag} by
including vector and t' Hooft interaction terms which can significantly
affect the equation of state of the compact stars with superconducting quark core~\cite{Klahn:2006iw,Bonanno:2011ch}. 
In this paper, our main aim is to investigate the competition between chiral and 
diquark condensates and therefore, we do not consider other interactions in our analysis.
 
We introduce auxiliary bosonic fields to bosonize the four-fermion interactions 
in Lagrangian~\eqref{eq:NJLlag} via a Hubbard-Stratonovich (HS) transformation.
The bosonic fields are 
\begin{eqnarray} 
\sigma = \left(\bar{q}q\right);~~\vec{\pi} = \left(\bar{q}i\gamma_5\vec{\tau}q\right);
~~\Delta = \left(\bar{q}^{C}i\gamma_{5}\epsilon_{f}\epsilon_{c}q\right);
~~\Delta^{*} =\left(\bar{q}i\gamma_{5}\epsilon_{f}\epsilon_{c}q^C\right);
\end{eqnarray}
and after the HS transformation, the bosonized Lagrangian density becomes
\begin{eqnarray}
\mathcal{L} &=& \bar{q}\left[i\gamma^{\mu}\left(\partial_{\mu} - ieQA_{\mu} - igT^8G^8_{\mu}\right) + \hat{\mu}\gamma^{0}\right]q 
- \bar{q}\left(m + i\gamma_5\vec{\pi}\cdot\vec{\tau}\right)q \nonumber\\
&-& \frac{1}{2}\Delta^{*}\left(\bar{q}^{C}i\gamma_{5}\epsilon_{f}\epsilon_{c}q\right) - \frac{1}{2}\Delta\left(\bar{q}i\gamma_{5}\epsilon_{f}\epsilon_{c}q^C\right)
- \frac{\sigma^2 +
\vec{\pi}^2}{4G_s} - \frac{\Delta^{*}\Delta}{4G_D}\ ,
\end{eqnarray}
where $m = m_0 + \sigma$. We set $\vec{\pi} = 0$ in our analysis, which excludes the possibility of pion condensation 
for simplicity \cite{Andersen:2007qv}.
Order parameters for chiral symmetry breaking and color superconductivity in the 2SC phase are
represented by non-vanishing vacuum expectation values (VEVs) for $\sigma$ and $\Delta$.
The diquark condensates of $u$ and $d$ quarks carry a net electromagnetic 
charge, implying that there is a Meissner effect for ordinary magnetism, while a linear 
combination of the photon and gluon leads to a ``rotated" massless $\mathrm{U}(1)$ field which 
is identified as the in-medium photon. We can write the Lagrangian in terms of 
rotated quantities using the following identity,
\begin{eqnarray}
\label{rotquan}
eQA_{\mu} + gT^8G^8_{\mu} = \tilde{e}\tilde{Q}\tilde{A}_{\mu} 
+ \tilde{g}\tilde{T}^8\tilde{G}^8_{\mu}\ .
\end{eqnarray}
In the {\it r.h.s.} of the Eq.~\eqref{rotquan} all quantities are rotated. In 
$flavor\otimes color$ space in units of the rotated charge of an electron 
$\tilde{e}={\sqrt{3}ge}/{\sqrt{3g^2+e^2}}$ the rotated charge matrix is
\begin{eqnarray}
\tilde{Q} = Q_f\otimes{\bf 1}_c - {\bf 1}_f\otimes \frac{T^8_c}{2\sqrt{3}}\ .
\end{eqnarray}
The other diagonal generator $T^3_c$ plays no role here because the degeneracy of color $1$ and $2$ 
ensures that there is no long range gluon $3$-field. 
We take a constant rotated background $\mathrm{U}(1)$ magnetic field ${\bf B} = B\hat{z}$
along $+z$ axis. The gapped 
2SC phase is $\tilde{Q}$-neutral, requiring a neutralizing background of strange quarks and/or electrons. 
The strange quark mass is assumed to be large enough at the moderate densities under consideration so that strange quarks do not play any dynamical role in the analysis.

\section{Thermodynamic potential}
\label{sec:therpot}

The partition function in the presence of an external magnetic field $B$ in the mean field 
approximation is given by  
\begin{eqnarray}
\mathcal{Z} = \mc{N}\int\lt[d\bar{q}\rt]\lt[dq\rt]\exp\lt\{\int_{0}^{\beta}d\tau\int d^{3}\vec{x}\lt(\tl{\mathcal{L}} - \frac{1}{2}B^2\rt)\rt\}\ ,
\end{eqnarray}
where $\mc{N}$ is the normalization factor, $\beta=T^{-1}$ is the inverse of the temperature 
$T$, $B$ is the external magnetic field and $\tl{\mathcal{L}}$ is the Lagrangian density in terms of the rotated quantities. The full partition function $\mc{Z}$ can be written as a
product of three parts, $\mc{Z}=\mc{Z}_c\mc{Z}_{1,2}\mc{Z}_3$. Here, $\mc{Z}_c$ serves as a constant multiplicative factor, $\mc{Z}_{1,2}$ denotes the contribution for quarks with color 
``1'' and ``2'' and $\mc{Z}_3$ is for quarks with color ``3''. These three parts can be expressed as
\begin{align}
\mc{Z}_c &= \mc{N}\exp\lt\{-\int_0^{\bt}d\tau\int d^3\vec{x}\lt[\frac{\sg^2}{4G_S}+\frac{\Dl^{2}}{4G_D}+\frac{B^{2}}{2}\rt]\rt\} \ , \\
\mc{Z}_{1,2} &= \int\lt[d\bar{Q}\rt]\lt[dQ\rt]\exp\lt\{\int_0^{\bt}d\tau\int d^3\vec{x}
\lt[\frac{1}{2}\mc{L}_\textrm{kin}\lt(Q,Q^{c}\rt)
\rt.\rt. \nn \\  
&+ \lt.\lt. \frac{1}{2}\tl{e}\tl{Q}\lt(\bar{Q}\kern+0.30em /\kern-0.70em {A}Q-\bar{Q}^c\kern+0.30em /\kern-0.70em {A}Q^c\rt) + \frac{1}{2}\bar{Q}\Dl^-Q^c+\frac{1}{2}\bar{Q}^c\Dl^+Q
\rt] \rt\} \ , \\
\mc{Z}_{3} &= \int\lt[d\bar{q}_3\rt]\lt[dq_3\rt]\exp\lt\{\int_0^{\bt}d\tau\int d^3\vec{x}\lt[\frac{1}{2}\mc{L}_\textrm{kin}\lt(q_{3},q_{3}^{c}\rt)
\rt.\rt. \nn \\  
&+ \lt.\lt. \frac{1}{2}\tl{e}\tl{Q}(\bar{q}_3\kern+0.30em /\kern-0.70em{A}q_3-\bar{q}_3^c \kern+0.30em /\kern-0.70em{A}q_3^c)\rt]\rt\}\ .
\end{align}
The kinetic operators $\mathcal{L}_{\textrm{kin}}\lt(q,q^{c}\rt)$ now read 
$(i\hskip -0.1cm\not\!\partial+\mu\gamma^0-M)$ where $M=m_0+\sg$ and we use the notation $\Delta^-(/\Delta^+)=-i\gamma_5\epsilon_{f}\epsilon_{c}\Delta(/\Delta^*)$. In $flavor \otimes color$ space in units of $\tilde{e}=\sqrt{3}ge/\sqrt{3g^2+e^2}$ the rotated charge matrix is given by $\tl{Q}=Q\otimes {\bf 1}_c - {\bf 1}_f\otimes T^8/2\sqrt{3}$. Here, ${\bf 1}_c$ and 
${\bf 1}_f$ are unit matrix on color and flavor spaces respectively. In our case, this translates to $\tilde{Q}$ charges $u_{1,2}=1/2, d_{1,2}=-1/2, u_3=1$ and $d_3=0$. With $s$-quarks as inert background, we also have $s_{1,2}=-1/2$ and $s_3=0$. Imposing the charge neutrality and 
$\beta$-equilibrium conditions is known to stress the pairing and lead to gluon condensation and a strong gluomagnetic field~\cite{Ferrer:2006ie}. The role of such effects has been studied 
in~\cite{Mandal:2012fq}, but here our focus is on the interdependence of the condensates and their response to the strong magnetic field. 

Evaluation of the partition function and the thermodynamic potential, $\Omega=-T\ln\mc{Z}/V$ 
(where $V$ is the volume of the system) is facilitated by introducing eight-component Nambu-Gorkov spinors for each color and flavor of quark, leading to 
\begin{eqnarray}
\ln Z_{1,2}=\frac{1}{2}{\rm ln}\lbrace{\rm Det}(\beta G^{-1})\rbrace ;\quad  {\rm ln}Z_{3}=\frac{1}{2}\ln \lbrace{\rm Det}(\beta G_0^{-1})\rbrace\,;\\ \nonumber
\end{eqnarray}
where $G$ and $G_0$ are the quark propagators and inverse of the propagators are given by
\begin{eqnarray}
G^{-1}=\bordermatrix{
&  & \cr
&[G_{0,\tilde{Q}}^+]^{-1} & \Delta^- \cr
&\Delta^+ & [G_{0,-\tilde{Q}}^-]^{-1} \cr}
\,,\quad G_0^{-1}=\bordermatrix{
&  & \cr
& [G_{0,\tilde{Q}}^+]^{-1} & 0\cr
&0 & [G_{0,-\tilde{Q}}^-]^{-1}\cr}\,,
\end{eqnarray}
with $[G_{0,\tilde{Q}}^{\pm}]^{-1}=(\hskip -0.1cm\not\!\partial\pm\mu\gamma^0+\tilde{e}\tilde{Q}\kern+0.30em /\kern-0.70em A-m)$. The determinant computation is simplified by re-expressing the $\tilde{Q}$-charges in terms of charge projectors in the color-flavor basis, following techniques applied for the CFL phase~\cite{Noronha:2007wg}. The color-flavor structure  of the condensates can be unraveled for the determinant computation by introducing energy projectors~\cite{Huang:2001yw} and moving to momentum space, whereby we find
\begin{eqnarray}
{\rm ln}Z_{1,2}&=&{\rm Tr}_{c,f}\sum_{a}\sum_{p_0, {\bf p}}[{\rm ln}(\beta^2(p_0^2-(E_{\Delta,a}^+)^2)\beta^2(p_0^2-(E_{\Delta,a}^-)^2))]\,,\nonumber\\
{\rm ln}Z_3&=&{\rm Tr}_{f}\sum_{a}\sum_{p_0, {\bf p}}[{\rm ln}(\beta^2(p_0^2-(E_{p,a}^+)^2)\beta^2(p_0^2-(E_{p,a}^-)^2))]\ ,
\end{eqnarray}
where $E_{\Delta,a}^{\pm}=\sqrt{(E_{p,a}^{\pm})^2+\Delta^2}$ with $E_{p,a}^{\pm}=E_{p,a}\pm\mu$ and $a=\{0,1,\pm 1/2\}$. The energy $E_{p,a}$ is defined as 
$E_{p,a}=\sqrt{{\bf p}_{\perp,a}^2+p_z^2+m^2}$, if $a=0$ then 
${\bf p}_{\perp,0}^2=p_x^2+p_y^2$ else ${\bf p}_{\perp,a}^2=2|a|\tilde{e}Bn$.
The sum over $p_0=i\omega_k$ denotes the discrete sum over the Matsubara frequencies, $n$ labels the Landau levels in the magnetic field which is taken in the $\hat{z}$ direction.

\section{Gap Equations and Solution}
\label{sec:gapeqn}

Using the following identity we can perform the discrete summation over the Matsubara frequencies
\begin{eqnarray}
\sum_{p_0}{\rm ln}[\beta^2(p_0^2-E^2)]=\beta[E+2T{\rm ln}(1+e^{-\beta E})\equiv\beta f(E)\,.
\end{eqnarray}
Then we go over to the 3-momentum continuum using the replacement
$\sum_{\bf p}\to V(2\pi)^{-3}\int d^3{\bf p}$, where 
$V$ is the thermal volume of the system.
Finally, the zero-field thermodynamic potential can be expressed as,
\begin{eqnarray}
\Omega_{B=0}=\frac{\sigma^{2}}{4G_{S}}+\frac{\Delta^{2}}{4G_{D}}-2\int_0^{\infty}\frac{d^3{\bf p}}{{(2\pi)}^3}[f(E_{p}^{+})+f(E_{p}^{-})+2f(E_{\Delta}^{+})+2f(E_{\Delta}^{+})]\,.
\end{eqnarray}
In presence of a quantizing magnetic field, discrete Landau levels suggest the following replacement
\begin{eqnarray}
\int_0^{\infty}\frac{d^3{\bf p}}{{(2\pi)}^3}\rightarrow\frac{|a|\tilde{e}B}{8\pi^{2}}\displaystyle\sum_{n=0}^{\infty}\alpha_{n}\int_{-\infty}^
{\infty}dp_{z}\ ,
\end{eqnarray}
where $\alpha_n=2-\delta_{n0}$ is the degeneracy factor of the $n$-th Landau level (all levels are doubly degenerate except the zeroth level). The thermodynamic potential in presence of magnetic field is given by
\begin{eqnarray}
\Omega_{B\neq 0}&=&\frac{\sigma^{2}}{4G_{S}}+\frac{\Delta^{2}}{4G_{D}}-\int_0^{\infty}\frac{d^3{\bf p}}{{(2\pi)}^3}[f(E_{p,0}^{+})+f(E_{p,0}^{-})]\nonumber\\
&-&\frac{\tilde{e}B}{8\pi^{2}}\displaystyle\sum_{n=0}^{\infty}\alpha_{n}\int_{-\infty}^
{\infty}dp_{z}[f(E_{p,1}^{+})+f(E_{p,1}^{-})+2f(E_{\Delta,\frac{1}{2}}^{+})+2f(E_{\Delta,\frac{1}{2}}^{-})]\ .
\end{eqnarray}

In either case, we can now solve the gap equations obtained by minimizing the (zero-temperature) thermodynamic potential $\Omega$ obtained in presence of magnetic field.
\begin{eqnarray}
\label{eq:gapeqn}
{\textrm {Chiral gap equation}}:~\frac{\partial\Omega}{\partial m}=0\,;\quad {\textrm {Diquark gap equation}}:~\frac{\partial\Omega}{\partial\Delta}=0.
\end{eqnarray}

Since the above equations involve integrals that diverge in the ultra-violet region, we must regularize in order to obtain physically meaningful results. We choose to regulate these functions using a sharp cut-off (step function in $|{\bf p}|$), which is common in effective theories such as the NJL model~\cite{Schwarz:1999dj,Buballa:2001gj}, although one may also employ a smooth regulator~\cite{Alford:1998mk,Noronha:2007wg} without changing the results qualitatively for fields that are not too large~\footnote{For example, a smooth cutoff was employed in~\cite{Noronha:2007wg} to demonstrate the De-Haas Van Alphen oscillations in the gap parameter at very large magnetic field.}. 
The momentum cut-off restrict the number of completely occupied Landau levels 
$n_{max}$ which can be determined as follows
\begin{eqnarray}
\int_0^{\Lambda}\frac{d^3{\bf p}}{{(2\pi)}^3}\rightarrow\frac{|a|\tilde{e}B}{8\pi^{2}}\displaystyle\sum_{n=0}^{n_{max}}\alpha_{n}\int_{-\Lambda^{'}}^
{\Lambda^{'}}dp_{z};~~n_{max}={\rm Int}\Bigr[\frac{\Lambda^2}{2|a|\tilde{e}B}\Bigl];~~\Lambda^{'}=\sqrt{\Lambda^{2}-2|a|\tilde{e}Bn}\,.
\end{eqnarray}
We use the fact that $p_{z}^{2}\geq 0$ to compute $n_{max}$. For magnetic field $B \lesssim 0.02$ {\rm GeV}$^2$ ($\sim 10^{17}$~G, conversion to Gauss is given by $1~{\rm GeV}^2=5.13\times 10^{19}$~G), $n_{max}$ is of the order of 50 and the discrete summation over Landau levels becomes almost continuous. In that case, we recover the results of the zero magnetic field case as described in the next section. For fixed values of the free parameters, we were able to solve the chiral and diquark gap equations self-consistently, for $B=0$ as well as large $B$. Before discussing our numerical results, we note the origin of the interdependence of the condensates. The chiral gap equation contains only $G_{S}$ which is determined by vacuum physics, but also depends indirectly on $G_D/G_S$ (a free parameter) through $\Delta$, which is itself dependent on the constituent $m=m_0+\sigma$. Our numerical results can be understood as a consequence of this coupling and the fact that a large magnetic field stresses the $\bar{q}q$ pair (same $\tilde{Q}$ charge, opposite spins implies anti-aligned magnetic moments) while strengthening the $qq$ pair (opposite $\tilde{Q}$ charge and opposite spins implies aligned magnetic moments).

\section{Numerical analysis}
\label{sec:nuana}

In order to investigate the competition between the chiral
and the diquark condensates, in this section, we solve the two coupled gap 
equations~\eqref{eq:gapeqn} numerically. These gap 
equations involve integrals that have diverging behavior in the high-energy region (this is an artifact of the
nonrenormalizable nature of the NJL model). Therefore, to 
obtain physically meaningful behavior, one has to regularize the diverging integrals by introducing some cutoff scale 
$\Lm$.
A sharp cutoff function sometimes leads to unphysical oscillations in thermodynamical quantities of interest, and especially for a system with discrete Landau levels. A novel regularization procedure called ``Magnetic Field
Independent Regularization'' (MFIR) scheme~\cite{Menezes:2008qt,Allen:2015paa} can remove the unphysical oscillations 
completely even if a sharp cutoff function is used within MFIR. To reduce the unphysical behavior, it is very common
in literature to use various smooth cutoff functions although they cannot completely remove the spurious
oscillations. Here, we list a few of them:
\begin{itemize}

\item Fermi-Dirac type~\cite{Fukushima:2007fc}:
$\displaystyle  f_c(p_a) = \frac{1}{2}\lt[ 1 - \tanh\lt( \frac{p_a - \Lm}{\al}\rt)\rt]$ where $\al$ is a smoothness 
parameter.

\item Woods-Saxon type~\cite{Fayazbakhsh:2010gc}: 
$\displaystyle f_c(p_a) =  \lt[ 1 + \exp\lt( \frac{p_a - \Lm}{\al}\rt)\rt]^{-1}$ where $\al$ is a smoothness 
parameter.

\item Lorenzian type~\cite{Frasca:2011zn}: 
$\displaystyle f_c(p_a) =  \lt[1+\lt(\frac{p_a^2}{\Lm^2}\rt)^N\rt]^{-1}$  where $N$ is a positive integer.
\end{itemize}
where $p_a = \sqrt{{\bf p}_{\perp,a}^2 + p_z^2}$, with ${\bf p}_{\perp,0}^2 = p_x^2 + p_y^2$ 
for $a=0$ and ${\bf p}_{\perp, a}^2=2|a|\tl{e}Bn$ for $a = 1,\pm 1/2$. Cutoff functions become smoother for larger values 
of $\alpha$, or $N$ in case of the Lorenzian type of regulator.
We have checked our numerical results for different cutoff schemes like sharp cutoff (Heaviside step function) and various
smooth cutoff parameterizations as mentioned above and found that our main results are almost insensitive for different cutoff schemes. We therefore,
use a smooth Fermi-Dirac type of regulator with $\al = 0.01\Lm$ throughout numerical analysis.

One can fix various NJL model parameters -- the bare quark mass $m_0$, the momentum cutoff $\Lambda$ and the scalar coupling constant $G_S$ by fitting the pion properties
in vacuum {\it viz.} the pion mass $m_{\pi}=134.98$ MeV, the pion decay constant $f_{\pi}=92.30$ MeV 
and the constituent quark mass $m(\mu = 0)=0.33$ GeV. Similarly, one can fix the diquark
coupling constant $G_D$ by fitting the scalar diquark mass ($\sim 600$ MeV) to obtain vacuum baryon mass of the
order of $\sim 900$ MeV~\cite{Ebert:1991pz}. There are some factors that can, in principle, alter those
model parameters {\it e.g.} strength of the external magnetic field, temperature, choice of
the cutoff functions etc. Assuming that those factors have only small effects on the 
parameters and expecting that our numerical results would not change qualitatively, we
fix the parameters in the isospin symmetric limit as follows (a discussion of the parameter choice
can be found in Ref.~\cite{Buballa:2003qv})
\begin{eqnarray}
\label{njlparam}
m_{u,d}=m_0=5.5~\textrm{MeV},~~\Lambda = 0.6533~\textrm{GeV},~~G_S = 5.0163~{\textrm{GeV}}^{-2},~~G_D=\rho G_S\ ,
\end{eqnarray}
where $\rho$ is a free parameter. Although Fierz transforming one gluon exchange
implies $\rho=0.75$ for $N_c=3$ and fitting the vacuum
baryon mass gives $\rho=2.26/3$ \cite{Ebert:1991pz}, the underlying interaction at moderate density is bound to be more complicated, therefore we choose to vary the coupling strength of the diquark channel $G_D$ to investigate the competition between the condensates.

\begin{figure}
\subfloat[]{\includegraphics[width=5cm,height=4cm]{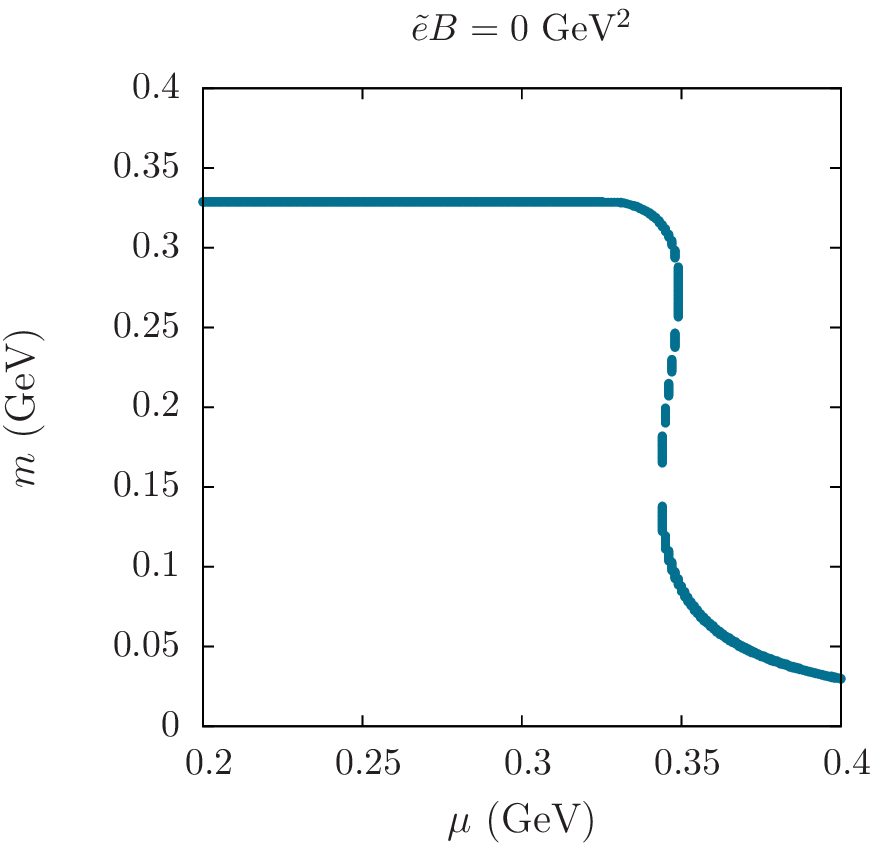}\label{fig:meB0del0}}
\subfloat[]{\includegraphics[width=5cm,height=4cm]{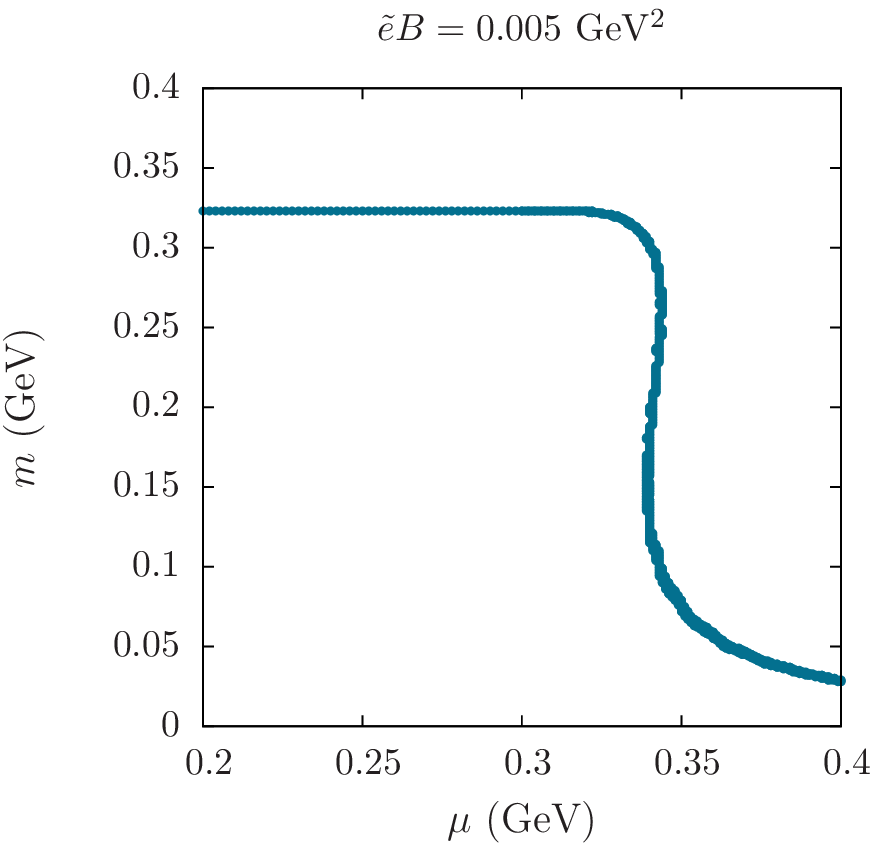}\label{fig:meB005del0}}
\subfloat[]{\includegraphics[width=5cm,height=4cm]{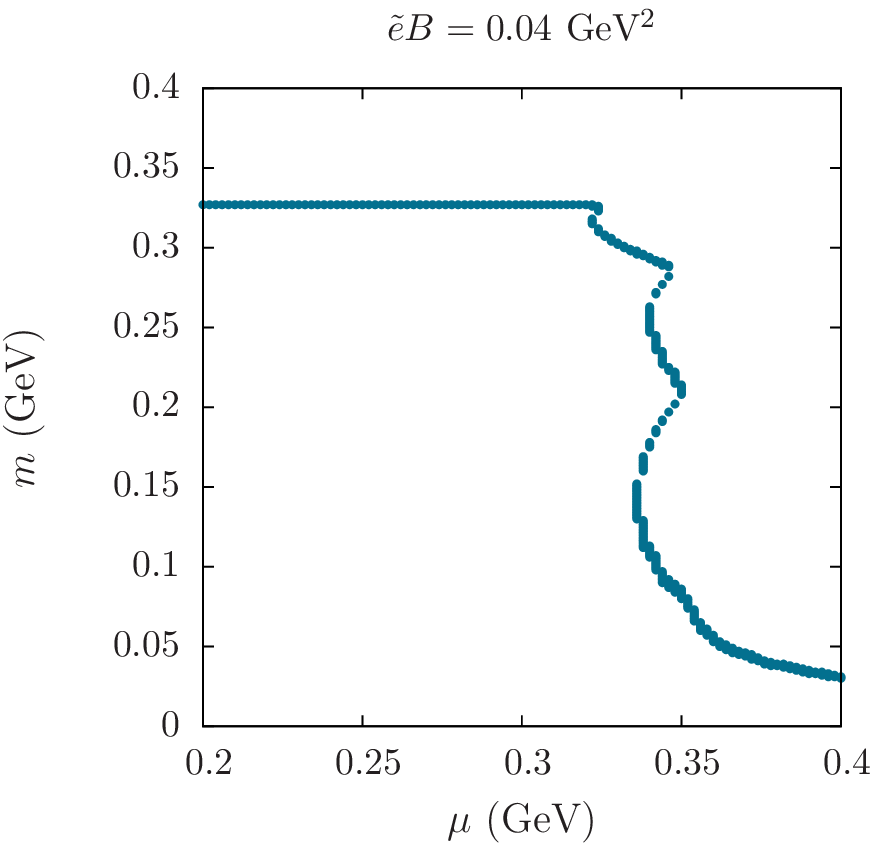}\label{fig:meB04del0}}\\
\subfloat[]{\includegraphics[width=5cm,height=4cm]{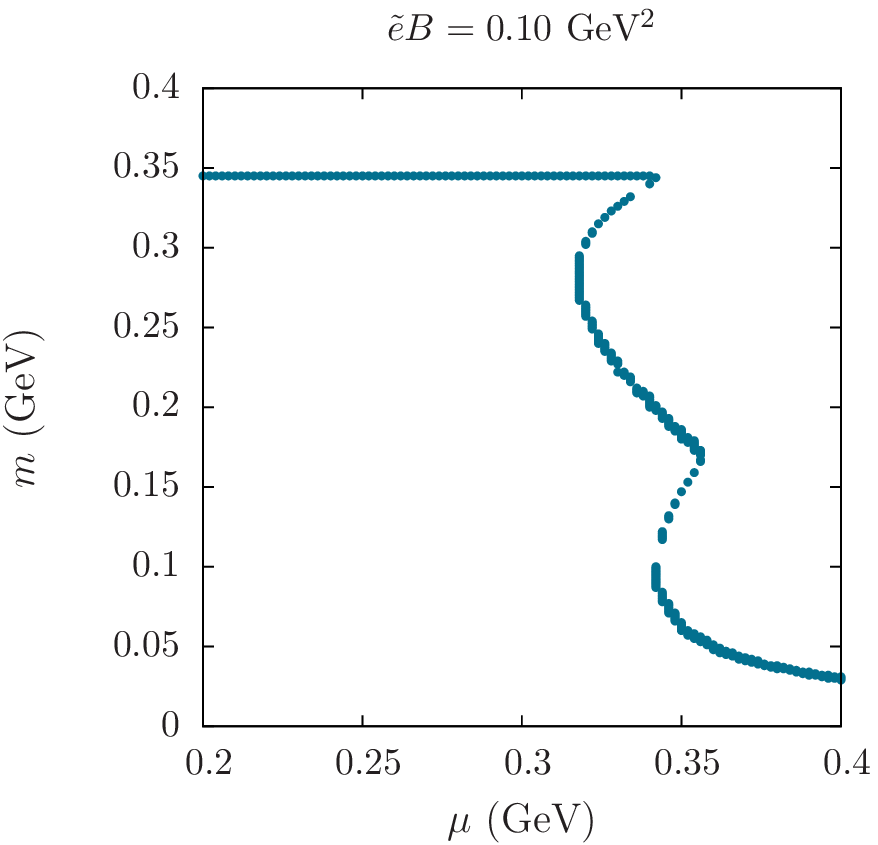}\label{fig:meB10del0}}
\subfloat[]{\includegraphics[width=5cm,height=4cm]{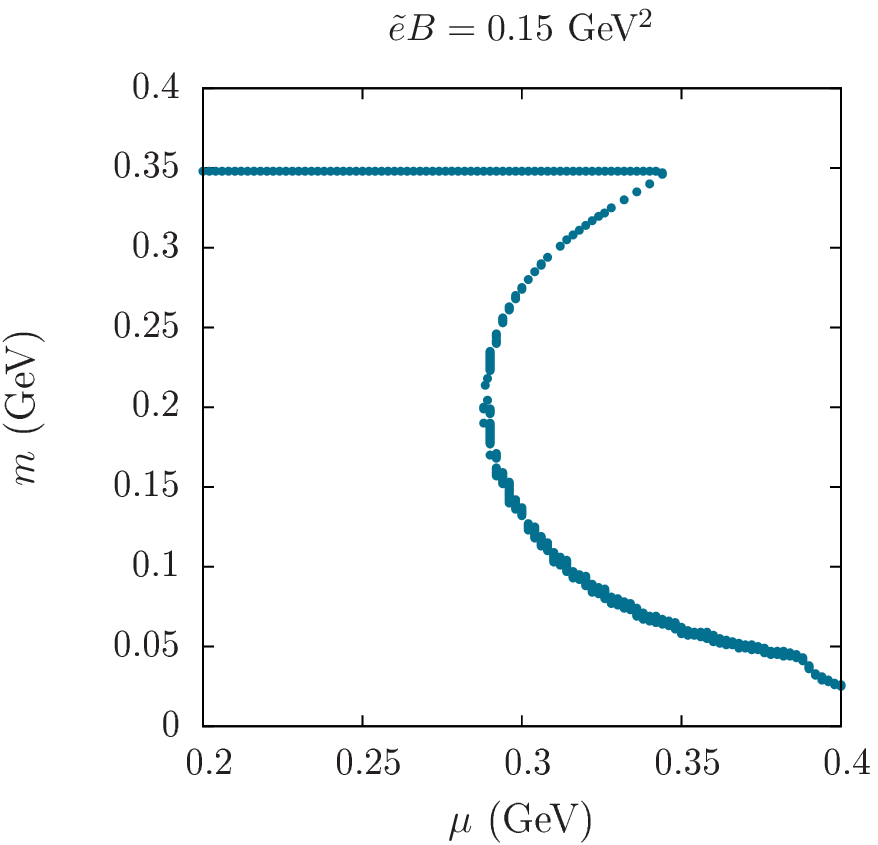}\label{fig:meB15del0}}
\subfloat[]{\includegraphics[width=5cm,height=4cm]{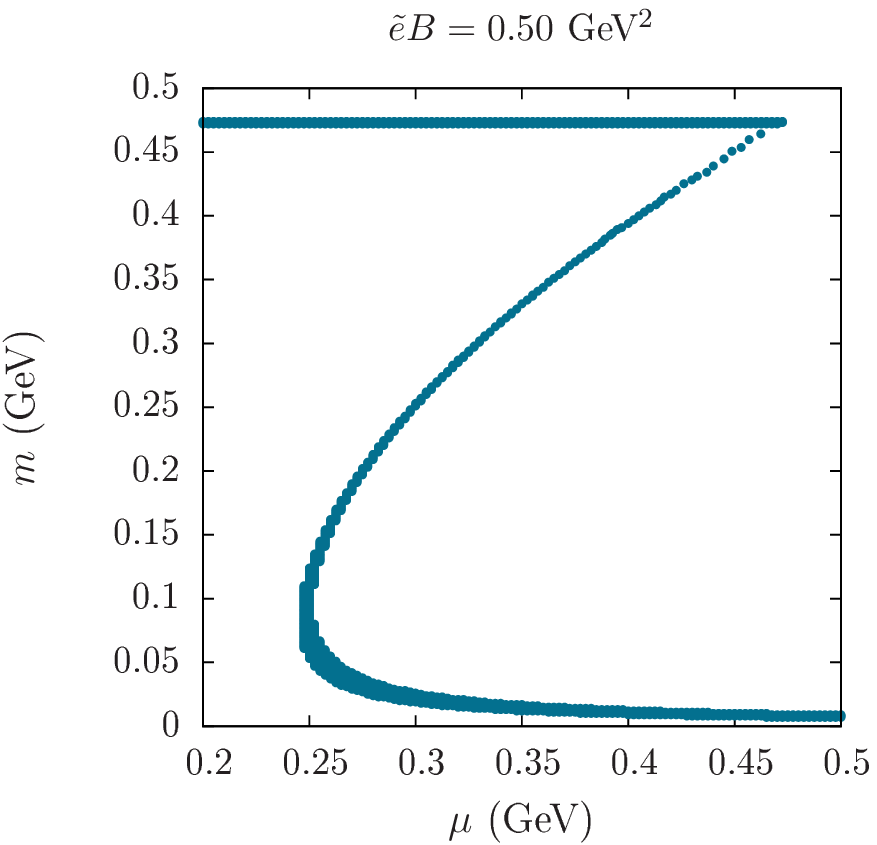}\label{fig:meB50del0}}
\caption{Chiral gap $m$ as functions of $\mu$ for $\rho=G_D/G_S=0$ with increasing magnetic field $\tl{e}B$ at $T=0$.}
\label{fig:meBdel0}
\end{figure}

We investigate the behavior of the chiral and diquark gaps along the
chemical potential direction in presence of magnetic field for different magnitudes
of the coupling ratio $\rho$ ($=G_D/G_S$) at zero temperature. Before we discuss the influence of diquark
gap on the chiral phase transition, we first demonstrate the behavior of the chiral
gap for $\Dl=0$ case (equivalently $\rho=0$) for different magnitudes of $\tl{e}B$.
The choice of $\tl{e}B$ is made to see the effects
of the inclusion of different Landau levels in the system. In Table~\ref{tab:LLeB}, we
show the values of $n^{max}_1$ and $n^{max}_{\frac{1}{2}}$ and the corresponding 
values of the transition magnetic field $\tl{e}B_t$. For example, if $\tl{e}B < \tl{e}B_t$, then
the number of fully occupied Landau level, $n > n^{max}$. In Fig.~\ref{fig:meBdel0}, we 
show $m$ as functions of $\mu$ in absence of diquark gap for different choices of $\tl{e}B$. 
In Figs.~\ref{fig:meB0del0} and~\ref{fig:meB005del0}, we show the $m$ in absence of magnetic
field ($\tl{e}B=0$) and in the weak magnetic field limit ($\tl{e}B=0.005$ GeV$^2$ or equivalently $\sim 2.5\times 10^{17}$~G) respectively. 
One can see that these two figures look almost identical. The reason is that the number of completely occupied
Landau levels, $n^{max}$ becomes very large ({\it e.g.} $n^{max}_1\sim 40$ for $\tl{e}B \sim 0.005$ GeV$^2$) in the weak magnetic field limit, making
the discrete Landau level summation quasi-continuous. As we increase the magnetic field,
noticeable deviations appear in the behavior of the chiral gap as seen in 
Figs.~\ref{fig:meB04del0} to \ref{fig:meB50del0}. 
\begin{center}
\begin{table}
\begin{tabular}{|c|c|c|c|}
\hline 
$n^{max}_1=$ & $\tl{e}B_t$ & $n^{max}_{\frac{1}{2}}=$ & $\tl{e}B_t$ \\ 
Int$\lt[\frac{\Lm^2}{2\tl{e}B}\rt]$ & (GeV$^2$) & Int$\lt[\frac{\Lm^2}{\tl{e}B}\rt]$ & (GeV$^2$) \\ 
\hline 
 1 & 0.213 & 1 & 0.427 \\
\hline 
 2 & 0.107 & 2 & 0.213 \\
\hline 
 3 & 0.071 & 3 & 0.142 \\
\hline 
 4 & 0.053 & 4 & 0.107 \\
\hline 
 5 & 0.043 & 5 & 0.085 \\ 
\hline
\end{tabular} 
\caption{The value of transition magnetic field $\tl{e}B_t$ at successive numbers of fully occupied
lowest Landau levels $n_a^{max}$ where $a=1,1/2$ are the two possible values of the rotated charge $\tl{Q}$.}
\label{tab:LLeB}
\end{table}
\end{center}

\begin{figure}
\subfloat[]{\includegraphics[width=7.8cm,height=7cm]{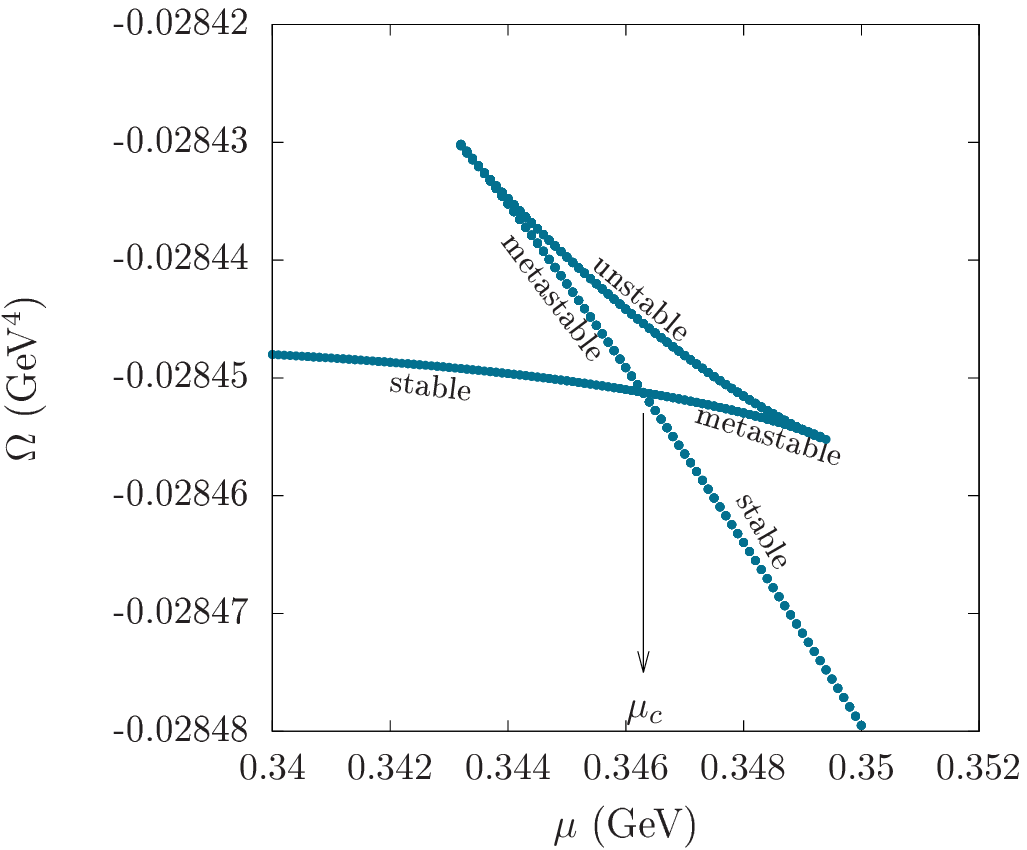}\label{fig:omgeB0}}
\subfloat[]{\includegraphics[width=7cm,height=7cm]{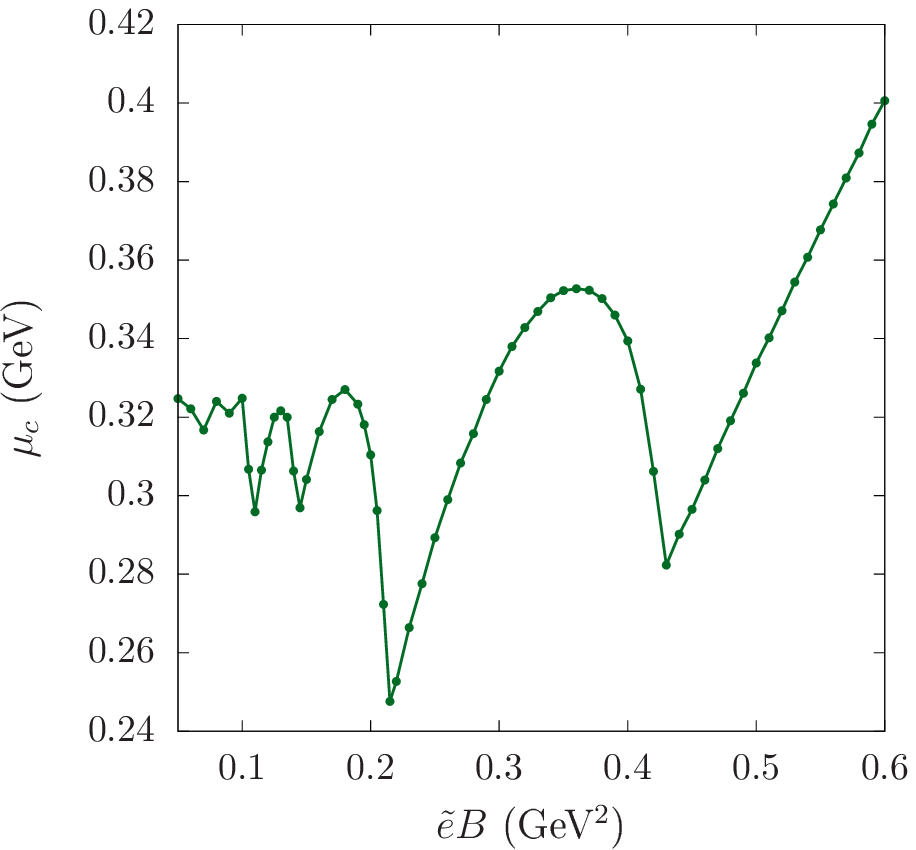}\label{fig:muF0}}
\caption{(a) Thermodynamic potential ($\Om$) for $\Dl=0$ and $\tl{e}B=0$ as a function of $\mu$. The stable
metastable and unstable branches of $\Om$ are shown beside the curves. (b) The critical chemical potential for chiral phase transition
($\mu_c$) for $\Dl=0$ as a function of $\tl{e}B$.}
\label{fig:omgmuF}
\end{figure}

From Fig.~\ref{fig:meBdel0}, it is clear that we get multiple solutions to the chiral gap equation for a small range of $\mu$
around the chiral phase transition region. For example, we get three solutions
to the chiral gap equation for a narrow window of $\mu$ for zero or weak magnetic field ($\lesssim 2.5\times 10^{17}$~G) cases. 
These three solutions correspond to the stable, metastable and unstable branches of the system. 
In Fig~\ref{fig:omgeB0}, we plot the values of $\Omega$ corresponding to the three 
solutions obtained in the small $\mu$-window. The value of the gap $m$ for which 
$\Omega$ is the lowest corresponds to the stable solution at any given density. 
The critical chemical potential $\mu_c$ (where the chiral and the diquark phase transitions occur)
is the point where the first derivative of $\Omega$ (and the gaps) behave discontinuously. 
The location of $\mu_c$ gives the transition point from the stable region to the metastable 
region of the system. This can easily be identified by looking at the behavior of $\Omega$
as shown in Fig.~\ref{fig:omgeB0}. We follow this method to locate
the first order phase transition point. In Fig.~\ref{fig:muF0}, we plot
$\mu_c$ as a function of $\tl{e}B$. We observe that $\mu_c$ oscillates 
with $\tl{e}B$ with dips
whenever $\Lambda^2/\left(2|a|\tilde{e}B\right)$ takes an integer value, following the Shubnikov de Haas-van Alphen 
effect. Similar oscillations in the density of states and various thermodynamic quantities are observed in metals 
in presence of magnetic field at very low temperature.
The magnitude of oscillations becomes more pronounced as we increase the magnetic field. If 
$\tl{e}B \gtrsim 0.21$ GeV$^2$ ($\sim 10^{19}$~G), only the zeroth Landau level is completely occupied as evident from
Table~\ref{tab:LLeB}. 

We observe multiple intermediate transitions (from Fig.~\ref{fig:meB04del0} to Fig.~\ref{fig:meB10del0})
due to the filling of successive Landau levels, and for a particular $\mu$, sometimes there
are two stable solutions at different densities for the
same pressure. Similar multiple solutions of the gaps has been observed  in the context of 
magnetized-NJL model with repulsive vector interactions~\cite{Denke:2013gha}.
Comparing the values of $m$ at $\mu=0$ for very strong magnetic field, one finds that 
$m$ increases with $\tl{e}B$. This is the magnetic catalysis effect~\cite{Klimenko:1990rh,Klimenko:1991he,Gusynin:1994re,Gusynin:1994xp}.
It is also interesting to note that with increasing magnetic fields, the 
spread of the metastable region (the $\mu$-window where we have multiple solutions) becomes wider. For example, the spread of the
metastable regions for $\tl{e}B=0.5$ GeV$^2$ is about 0.22 GeV and for $\tl{e}B=0.25$ GeV$^2$ is 
about 0.12 GeV. These findings suggest the possibility of multiple phases with different values of dynamical mass in the presence of inhomogeneous magnetic fields, which we postpone to a future investigation.  
It is important to mention that the multiple solutions observed in chiral condensate
as function of chemical potential would disappear when plotted as function of baryon
density defined as $\langle\bar{q}\gm^0q\rangle$ (see e.g. Section 4.2 in Ref.~\cite{Schwarz:2003}).

\begin{figure}
\subfloat[]{\includegraphics[width=7.4cm,height=7.3cm]{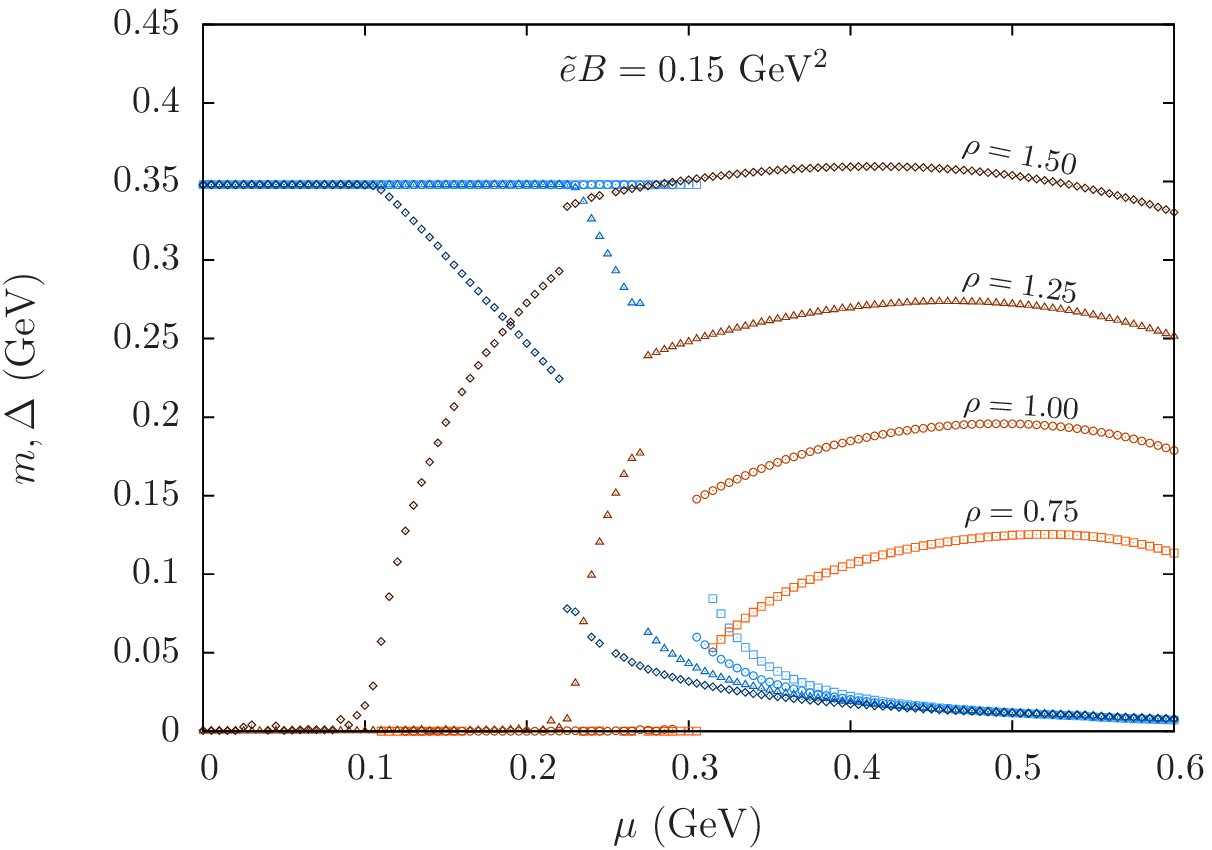}\label{fig:mdeB50}}
\subfloat[]{\includegraphics[width=7.2cm,height=7.2cm]{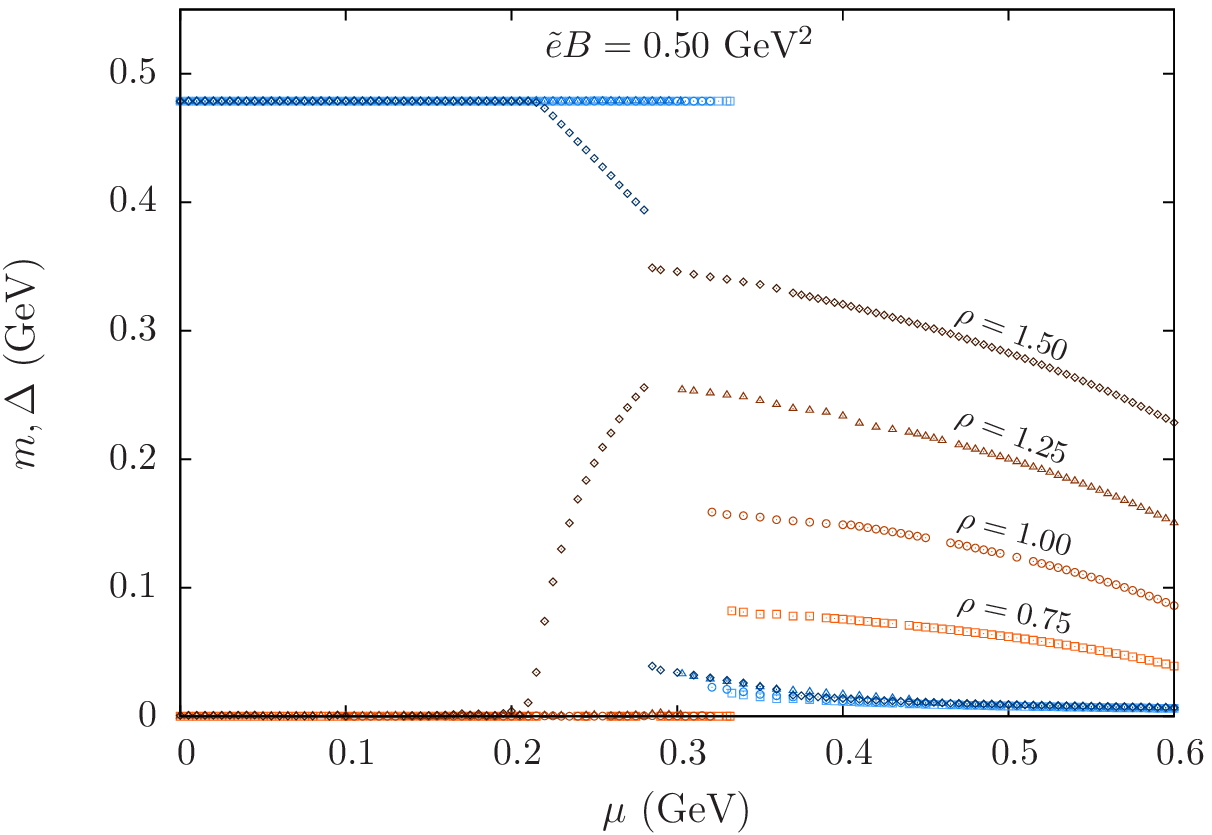}\label{fig:mdeB15}}
\caption{The gaps $m$ (different blue colors) and $\Dl$ (different brown colors) as functions of $\mu$ for different $\rho$ in presence of strong magnetic field (a) $\tl{e}B=0.15$ GeV$^2$ and (b) $\tl{e}B=0.5$ GeV$^2$.
The curves with square, circle, triangle and diamond represent $\rho=0.75,1,1.25,1.5$ respectively.
The discontinuities in gaps signify a first-order phase transition.}
\label{fig:mdmu}
\end{figure}

In Fig.~\ref{fig:mdmu}, we show $m$ and $\Dl$ as functions of $\mu$ for different $\rho$ in presence of 
strong magnetic field. In \cite{Huang:2001yw}, the competition of chiral and
diquark gaps without any magnetic field was discussed in great detail. We observe that $m$ increases with the 
increase of $\tl{e}B$. For example, $m\sim 0.35$ GeV for $\tl{e}B=0.15$ GeV$^2$ and $m\sim 0.48$ GeV for $\tl{e}B=0.5$ GeV$^2$.
In \cite{Huang:2001yw}, it was shown that with the increase of $\rho$, the first order transition 
of the chiral and
diquark gaps becomes crossover through a second-order phase transition. When a strong magnetic field is present, we find that the crossover becomes a first order transition. This is an important finding of this work, which has several implication for neutron star physics as discussed in the conclusion.
The critical chemical potential $\mu_c$ is almost same for both the chiral and diquark phase
transition, but takes on smaller values as we increase $\rho$ for a fixed $\tl{e}B$ (as shown in Fig.~\ref{fig:mdmu}).

In the weak (or zero) magnetic field limit, $\Dl$ appears at a smaller $\mu$ with increasing $\rho$
and rises smoothly from zero, until it becomes discontinuous at $\mu_c$. At $\mu_c$, the chiral gap $m$
also changes discontinuously, with the jumps in the gaps decreasing with increasing $\rho$. For instance, in Table~\ref{tab:jump} we see
the jump in the chiral gap ($\delta_m$) decreases from 0.250 GeV to 0.133 GeV as we increase $\rho$ from 0.75 to 1.05.
The corresponding jumps in diquark gap ($\delta_{\Dl}$) decreases from 0.077 GeV to 0.066 GeV. This picture does not change qualitatively until we go above a 
critical value $\rho=\rho_c \approx 1.09$. As long as $\rho < \rho_c$, the jumps in the gaps $\delta_m$ and $\delta_{\Delta}$ 
remain nonzero but decrease as $\rho$ moves towards $\rho_c$. In other words, the metastable region in $\Omega$ as shown in Fig.~\ref{fig:omgeB0}
shrinks with increasing $\rho$. At $\rho=\rho_c$ the metastable and unstable regions vanish completely. 
This qualifies it to be a second order phase transition. Above $\rho > \rho_c$, the gaps $m$ and $\Delta$ 
are smoothly varying resulting in a smooth crossover. 
However, there always exists a pseudo-transition point ($\mu_c^p$) around which fluctuations/variations 
of the condensates ({\it i.e.} derivatives of $m$ and $\Dl$ {\it w.r.t.} $\mu$) are sharply peaked. 
The width of these peaked distributions broaden with further increase of $\rho$, 
and $\mu_c^p$ moves to the left with increasing $\rho$. 
These results for $\tl{e}B\approx 0$ agree qualitatively 
with the zero field results of \cite{Huang:2001yw} with minor quantitative differences at less than a few percent level.
The region where the condensates 
coexist was termed by them as the ``mixed broken phase", since both chiral and (global) color symmetries
are broken here. While it should not be confused with a genuine mixed phase, since the free energy admits a unique 
solution to the gap equations in this regime, it is clear that the width of this overlap region increases with increasing $\rho$. 

\begin{table}
\begin{tabular}{|c|c|c|c|c|c|c|c|c|}
\hline
$\rho$ & \multicolumn{4}{|c|}{$\tl{e}B=0$} & \multicolumn{4}{|c|}{$\tl{e}B=0.1$ GeV$^2$} \\
\cline{2-9}
& $\delta_m$ (GeV) & $\delta_{\Delta}$ (GeV) & $\mu_c$ (GeV) & Nature & $\delta_m$ (GeV) & $\delta_{\Delta}$ (GeV) & $\mu_{c}$ (GeV) & Nature \\
\hline
0.75 & 0.250 & 0.077 & 0.332 & First order & 0.227 & 0.084 & 0.323 & First order \\
\hline
1.05 & 0.133 & 0.063 & 0.295 & First order & 0.214 & 0.074 & 0.301 & First order \\
\hline
1.09 &    0  &     0 & 0.289 & Second order& 0.198 & 0.066 & 0.297 & First order\\
\hline
1.15 & Smooth & Smooth & 0.280 & Crossover & 0.185 & 0.051 & 0.295 & First order\\
\hline
1.25 & Smooth & Smooth & 0.255 & Crossover & 0.122 & 0.038 & 0.284 & First order\\
\hline
\end{tabular}
\caption{Critical chemical potential $\mu_c$, jumps in the chiral ($\delta_m$) and superconducting ($\delta_{\Delta}$) order parameters at $\mu_c$ for zero and 
$\tl{e}B=0.1$ GeV$^2$ ($\sim 5\times 10^{18}$~G) for various values of $\rho=G_D/G_S$. The nature of the transition is also indicated.}
\label{tab:jump} 
\end{table}

The competition between the condensates is driven by the strong magnetic field, which in the case of $m$ is a stress, since the chiral condensate involves quark spinors of opposite spin and same $\tilde{Q}$-charge. On the other hand, the diquark condensate, with opposite spin and $\tilde{Q}$-charge, 
is strengthened by the strong magnetic field. Thus, we expect a strengthening of the competition between the two condensates, 
resulting in a qualitative change from the zero-field case. 
With increasing $\rho$, similar to the $\tl{e}B=0$ case, $\delta_m$ and $\delta_{\Delta}$ decrease and the transition is
first order in nature. The dramatic effect
we observe is that the mixed broken phase for large $\rho$ at $\tl{e}B=0$ is no longer present in case of strong magnetic field case
and the crossover region is replaced by a first-order transition. Specifically, in Table~\ref{tab:jump}, we see for $\rho=1.25$, a smooth crossover in the $\tl{e}B=0$ case at $\mu_c^p \sim 0.255$ GeV becomes a first order transition with $\delta_m = 0.185$ GeV and $\delta_{\Dl}=0.122$ GeV at $\mu_c \sim 0.284$ GeV for $\tl{e}B=0.1$ GeV$^2$.
The simultaneous appearance of the discontinuity in the gaps for large magnetic field case, at almost the same $\mu=\mu_c$ where
both the condensates have their most rapid variation in the $\tl{e}B=0$ case, is a physical feature and is also cutoff insensitive. 
We have checked that magnetic field $\tl{e}B \lesssim 5\times 10^{17}$~G does not notably alter the competition between 
the condensates from the zero magnetic field case.

\section{Conclusions}
\label{sec:conclu}

We study the effects of a strong homogeneous magnetic field on the chiral and
diquark condensates in a two-flavor superconductor using the NJL model.
We implement a self-consistent scheme to determine the condensates,
by numerically iterating the coupled (integral) equations for the chiral
and superconducting gaps. We obtain results for the nature of the competition 
between these condensates in two cases, at weak magnetic field limit where
our results are qualitatively same as zero magnetic field results~\cite{Huang:2001yw}
and at strong magnetic field, where we find the competition between the gaps increases 
strongly causing a discontinuity in the gaps and disrupting the ``mixed broken phase''. 
This is a result of the 
modified free energy of the quarks in the condensate when subjected to a strong
magnetic field. For magnetic fields as large as $B\sim 10^{19}$ G, the 
anti-aligned magnetic moments of the quarks in the chiral
condensate change the smooth crossover of the chiral transition to a sharp 
first order transition. The diquark gap also becomes discontinuous at this 
point. For magnetic fields $B \lesssim 10^{18}$ G, there is no significant effect of the magnetic field on the competition between the condensates and zero-field results apply.

These findings can impact the physics of hybrid stars (neutron stars with quark matter) or strange quark stars in several ways. 
Firstly, the structure of neutron stars is strongly affected by a first-order phase transition, with the possibility of a third family of compact stars in addition to neutron stars and white 
dwarfs~\cite{Ayriyan:2015kit} that is separated from conventional neutron stars by a radius gap of a few km. We can speculate that strange stars or hybrid stars with superconducting quark cores inside them belong to this third family. Since we find that a strong magnetic field increases the likelihood of a first-order phase transition and hence a mixed phase,  magnetars could also possibly belong to this category of compact stars since they permit quark nucleation~\cite{Kroff:2014qxa} and carry large interior magnetic fields which modify their mass-radius relationship~\cite{Orsaria:2010xx}.
Secondly, it was pointed out in~\cite{Mandal:2012fq} that for large values of the local magnetic field and in the small density
window of the metastable region, it is possible to realize domains or 
nuggets of superconducting regions with different values for the gap. Charge neutrality can also
disrupt the mixed broken phase, but the oscillations of the chiral gap remain, leading to nucleation of chirally restored 
droplets. Such kinds of nucleation and domain formation 
will release latent heat that might be very large owing to the large value of the magnetic field, 
serving as an internal engine for possible energetic events on the surface of the neutron 
star~\cite{Zdunik:2007dy,Mallick:2012zq}. Such internal mechanisms are unlikely to occur in a pure neutron star without a quark 
core. Thirdly, strong magnetic fields and quark cores affect the radial and non-radial oscillation modes of neutron stars, which could be a discriminating feature in the gravitational wave signal from vibrating neutron stars. The frequency of the fundamental radial mode shows a kink at the density characterizing the onset of the mixed phase, and the frequencies depend on the magnetic field~\cite{Panda:2015zxa}. Non-radial modes such as $g$-modes can probe the density discontinuity arising as a result of the phase transition in neutron stars~\cite{Miniutti:2002bh} or strange quark 
stars~\cite{Sotani:2001bb}, although the effect of magnetic fields in this context is as yet unexplored. 
Another important aspect of rotating compact stars are the $r$-modes~\cite{Andersson:1997xt}, which could be responsible for spinning down neutron stars or strange quark stars from their Kepler frequency down to the observed values seen in 
low-mass X-ray binaries. The effects of a strong magnetic field on the $r$-mode driven spin down of neutron stars have been studied in~\cite{Huang:2009ue,Staff:2011zn}, while $r$-modes in crystalline quark matter are discussed in~\cite{Knippel:2009st}. 
The even-parity counterpart for the $r$-modes, which include non-radial oscillation modes such as the $f$- and $p$-modes have also been explored for the case of strange quark stars in~\cite{Sotani:2003zc,Flores:2013yqa}.
Our findings give additional motivation to the study of such interesting effects associated with a first-order transition in neutron stars with strong magnetic fields, and a systematic study of these effects in the new era of gravitational waves and neutron star observations may finally reveal the presence of quark matter in the core of neutron stars.

\section*{Acknowledgments}

T.M. is supported by funding from the Carl Trygger Foundation under contract CTS-14:206 and 
the Swedish Research Council under contract 621-2011-5107. P.J. is supported by the National Science Foundation under Grant No. PHY 1608959.

\bibliography{m2SC}{}
\bibliographystyle{unsrt}

\end{document}